\documentclass{svjour3}                     % onecolumn (standard format)
\smartqed  % flush right qed marks, e.g. at end of proof
\usepackage{graphicx}
\usepackage{amsmath}
\usepackage{amsfonts}
\usepackage{amssymb}
\usepackage{bm}
\usepackage{subfig}
\newcommand{\lea}{\raisebox{-.3ex}{\small $ \
\stackrel{\textstyle <}{\sim} $ }}

\newcommand{\beq}{\begin{equation}}
\newcommand{\eeq}{\end{equation}}
\newcommand{\beqa}{\begin{eqnarray}}
\newcommand{\eeqa}{\end{eqnarray}}

\begin{document}

\bibliographystyle{unsrt}

\title{What is {\it ab initio}?}
%Grants or other notes


\titlerunning{What is {\it ab initio}?}        % if too long for running head

\author{R. Machleidt        
%\and
%        Second Author %etc.
}


\institute{R. Machleidt \at
              Department of Physics, University of Idaho, Moscow, ID 83844, USA\\
              \email{machleid@uidaho.edu}           %  \\
}

\date{Received: date / Accepted: date}

\maketitle

\begin{abstract}
Microscopic nuclear theory is based on the tenet
that atomic nuclei
can be accurately described as collections of point-like nucleons interacting via
two- and many-body forces obeying nonrelativistic quantum 
mechanics---and the concept of the {\it ab initio} approach is to calculate
nuclei accordingly.
The forces are fixed in free-space scattering and must be accurate.
We will critically review the history of this approach from the early beginnings until today.
An analysis of
current {\it ab initio} calculations reveals that some mistakes
of history are being repeated today.
The ultimate goal of nuclear theory are high-precision {\it ab initio}
calculations which, as it turns out, may be possible only at the fifths order of the chiral
expansion. Thus, for its fulfillment, nuclear theory is still facing an enormous task.
\end{abstract}

\section{Introduction}

The tenet of microscopic nuclear theory is that atomic nuclei
can be accurately described as collections of point-like nucleons interacting via
two- and many-body forces obeying nonrelativistic quantum 
mechanics---the forces being fixed in free-space scattering.

The microscopic or {\it ab initio} approach to nuclear structure
and reactions is then defined as calculating the properties of nuclei
in accordance with the tenet.

It is the purpose of this note to discuss how consistent or inconsistent
 the fundamental model of nuclear theory has been pursued through
 the history of nuclear physics and to provide an outlook for the future.

\section{Early history of the microscopic approach}

The microscopic approach to nuclear structure is almost as old as 
nuclear physics itself. Brueckner and co-workers introduced Brueckner theory as early as 1954~\cite{BLM54} and performed the first semi-realistic microscopic nuclear matter 
calculation in 1958~\cite{BG58}. 
Already that same year, Brueckner discussed  finite
nuclei proposing  the local
density approximation~\cite{BGW58}.

In the second half of the 1960's, 
one of the hottest topics in nuclear structure physics was calculating
the properties of
finite nuclei without recourse through nuclear matter using 
Brueckner-Hartree-Fock (BHF) theory.
The Oak Ridge National
Laboratory (ORNL) with its computer power played a leading role in this effort that was guided by Thomas Davies and Michel Baranger~\cite{Bar69,Dav69}.
BHF (and coupled cluster) calculations of finite nuclei continued into the early 1970s with work by the Bochum~\cite{KZ73} and the Bonn-J\"ulich groups~\cite{MMF75}.

In parallel to the above developments, research on the microscopic derivation of the shell-model effective interaction was conducted (again, applying Brueckner theory) that had been kicked off by Kuo and Brown in 1966~\cite{KB66}.

Applying the nucleon-nucleon ($NN$) potentials available at the time, 
the BHF approach reproduced about one half of the binding energies of closed-shell nuclei which, in the early phase, was seen
as a great success~\cite{Bar69}, but in the long run did not satisfy demands for
more quantitative predictions.
Therefore, a departure from the microscopic approach happened around 1973
as reflected most notably in a lead-talk by Michel Baranger at the
 International Conference on Nuclear Physics in Munich in 1973~\cite{Bar73}.
 
 The shell-model effective interaction suffered a similar fate at the
International Conference on Effective Interactions 
and Operators in Nuclei in Tucson, Arizona, in 1975, organized
by Bruce Barrett~\cite{Bar75}.

And so it happened that in the early 1970s, the microscopic approach was abandoned and replaced by phenomenological
effective interactions (also know as mean-field models):
the Skyme interaction~\cite{Sky59} as revived by Vautherin and co-workers~\cite{VB72,Vau73}, the Gogny force~\cite{Gog73,DG80}, and the relativistic  
mean-field model of Walecka~\cite{Wal74,SW86}.

Ironically, the calculations with those effective interactions continued to be called ``microscopic'', for which John Negele had provided
the (debatable) justification in his Ph.D.\  thesis of 1970~\cite{Neg70}. Before calculating finite nuclei in the local density approximation, Negele had adjusted the insufficient binding of nuclear matter provided by the Reid soft-core
potential~\cite{Rei68} (11 MeV per nucleon) by hand to the presumed empirical value
of 15.68 MeV making ``the assumption that when higher-order corrections have been evaluated carefully, nuclear-matter theory will indeed produce the
correct binding''~\cite{Neg70}.
Negele had many followers~\cite{CS72,FN73,MHN75}.

However, the true ``deeper reason'' for those effective interactions was 
much simpler:
``To get better results!''~\cite{ano70}.
Clearly, the trends
that won popularity in the early 1970s were a setback for the fundamental
research in nuclear structure.

Nuclear structure theory at its basic level is not about fitting data
to get ``good'' results.
Fundamental nuclear structure theory is about answering the question:

\begin{quote}
{\it Do the same nuclear forces that explain free-space scattering experiments
also explain the properties of finite nuclei and nuclear matter when
applied in nuclear many-body theory?}
\end{quote}

One can think of many reasons why the basic tenet should be wrong.
According to the EMC effect, nucleons swell when inserted into nuclei which
might affect the force between nucleons~\cite{Ban92}. Meson exchange in the nuclear medium may be different than in free-space
for various reasons~\cite{KMS76,Wil79,BKR91}. The excitation of resonances, e.~g.\
$\Delta(1232)$ isobars, within the nucleon-nucleon interaction process is subject to changes when 
happening in a nuclear medium~\cite{GN75,Gre76,DC76,HM77}.
And many more ideas have been advanced, like e.~g., Brown-Rho scaling~\cite{BR91}.
In fact, in the 1970s, a popular belief was that medium effects on the $NN$
interaction may be the solution to the problem of lacking saturation~\cite{Mac89}.

Thus, it is a good question to ask whether medium modifications 
of nuclear forces show up in a noticeable way and/or are even needed for quantitative nuclear structure predictions.
But when we re-adjust the free-space forces arbitrarily to get ``good'' results, then we will never find out.
Note also that at some (high) energy and high density, the picture of
point-like nucleons is bound to break down~\cite{Ben23}. 
So, the issue behind the nuclear theory tenet is:
 Are the energies typically involved in conventional nuclear structure physics low enough to treat nucleons as structure-less objects?

To come back to history: the renunciation of the truly microscopic approach
lasted about two decades (essentially the 1970s and 80s).
Then, in the early 1990s, the microscopic
theory was revived by the Argonne-Urbana group~\cite{Wir93,Pud95}.
The crucial element in those new microscopic calculations was the
inclusion of a three-nucleon force (3NF).
The idea of a nuclear 3NF was not new. In fact, it is almost as old as 
meson theory itself~\cite{PH39}. But for years it had been considered
just an academic topic, too difficult to incorporate into actual
calculations, anyhow. But the persistent failure to saturate nuclear matter
at reasonable energies and densities, as well as the the underbinding of
nuclei, finally compelled nuclear structure physicists
to take a serious look at the 3NF issue, as explained in the
exemplary Comment by Ben Day~\cite{Day83} based upon
first test calculations by the Urbana group~\cite{CPW83}.
The 3NF definitely improved nuclear saturation and the properties
of light nuclei, even though nothing was perfect~\cite{Pud95}.

\section{Recent history}

After the year of 2000, two changes occurred.
First, the term `microscopic' was increasingly
replaced by the term {\it `ab initio'}~\cite{NVB00}---for reasons
nobody knows (but nothing to worry about because both mean the same).
Second and more importantly, nuclear forces based upon chiral effective field theory (EFT) entered the picture~\cite{EGM00,EM03}. This development was of great advantage.
Note that for a microscopic approach to be truly microscopic, the free-space forces need to be accurate.
But with phenomenological or meson-theoretic forces it was difficult to define what sufficiently accurate means, since the errors in those theories are unknown. 
However, in the framework of an EFT, the theoretical uncertainty can
be determined and, thus, related with the accuracy of the predictions.
Hence, in the framework of an EFT:

\begin{quote}
{\it  Accurate free-space forces are forces that
predict experiment within the theoretical uncertainty of the EFT at the 
given order.}
\end{quote}

After 2000, it also became well established that predictive nuclear structure must include 
3NFs, besides the usual two-nucleon force (2NF) contribution.
Another 
advantage of chiral EFT is then that it generates 2NFs and multi-nucleon forces simultaneously and 
on an equal footing. In the $\Delta$-less theory~\cite{ME11,EHM09}, 3NFs occur for the first time at next-to-next-to-leading order
(NNLO) and continue to have additional contributions in higher orders.
 If an explicit
$\Delta$-isobar is included in chiral EFT ($\Delta$-full theory~\cite{ORK96,KGW98,KEM07,EKM08}), then 3NF contributions start already at next-to-leading order (NLO). 

In the initial phase, 
the 3NFs were typically adjusted in
 $A=3$ and/or the $A=4$ systems and the
 {\it ab initio} calculations were driven up to the oxygen region~\cite{BNV13}.
 It turned out that for $A \lea 16$ the ground-state energies and radii are predicted about right, no matter what type of chiral or phenomenological potentials were applied (local, nonlocal, soft, hard, etc.)
 and what the details of the 3NF adjustments to few-body systems were~\cite{BNV13,Rot11,Pia18,Lon18,Mar21,Mar22}.

 However, around the year of 2015, the picture changed, when the many-body practitioners were able to move up to medium-mass nuclei (e.~g., the calcium or even the tin regions). 
 Large variations of the predictions now occurred depending on what forces were used, and cases
 of severe underbinding~\cite{Lon17} as well as of substantial overbinding~\cite{Bin14} were observed. Ever since, the nuclear structure community understands that 
 accurate {\it ab initio} explanations of intermediate and heavy nuclei is an outstanding problem.
 
  There have been several attempts to 
  predict the properties of
  medium-mass nuclei with more accuracy.
Of the various efforts, we will now list four cases, which are representative
for the status, and will denote each case with a short label for
ease of communication. We restrict ourselves to cases, where 
the properties of medium-mass nuclei {\it and}\/ nuclear matter have been calculated, because the simultaneous description of both systems is part of the 
problem.\footnote{Other interesting cases are the models by
Soma {\it et al.}~\cite{Som20} and Maris {\it et al.}~\cite{Mar22}
for which, however, presently no nuclear matter results are available.}

 \begin{figure}[t]\centering
%\vspace*{-4.4cm}
\scalebox{0.60}{\includegraphics{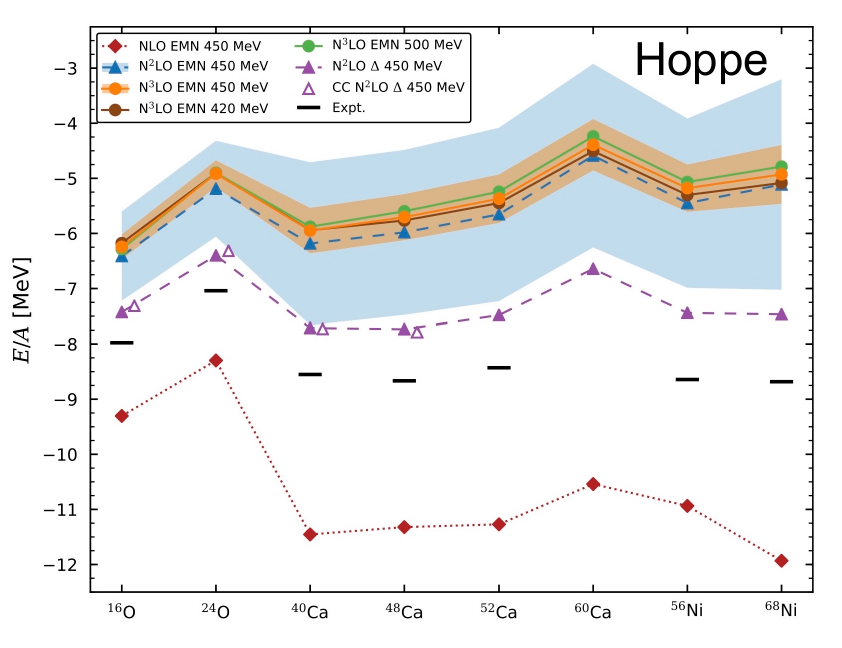}}

\scalebox{0.60}{\includegraphics{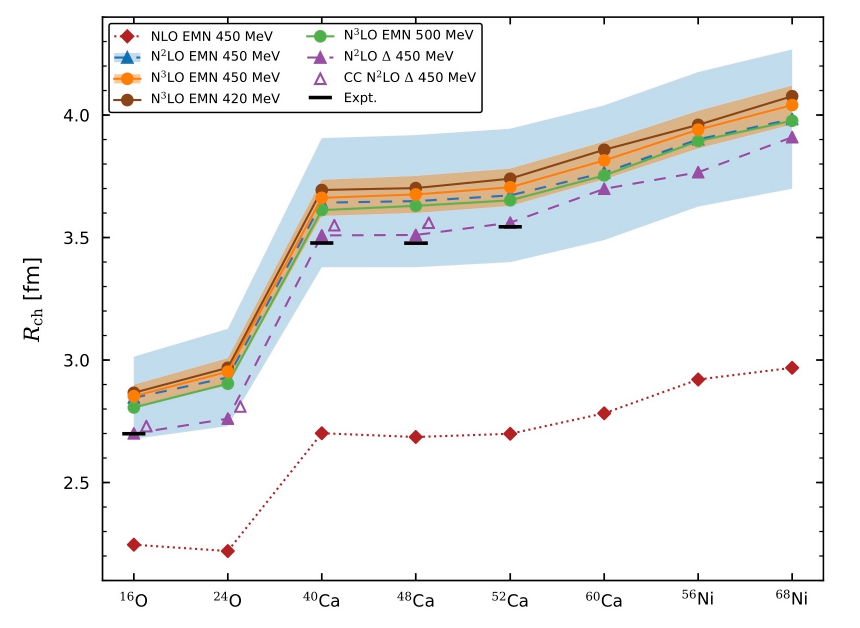}}

%\vspace*{-0.25cm}
\caption{Upper panel:
Ground-state energies per nucleon, $E/A$, of selected closed-shell oxygen, calcium, and nickel isotopes 
as obtained in the 
 ``Hoppe'' case~\cite{Hop19}.
Results are shown for various chiral interactions as denoted.
 The blue and orange bands give the NNLO and N$^3$LO uncertainty 
 estimates, respectively. 
 $\Lambda=450$ MeV in all cases except the green curve.
 Black bars indicate experimental data.
   Lower panel: 
 Same as upper panel, but for charge radii.
 (Reproduced from Ref.~\cite{Hop19} with permission.)
  }
 \label{fig_hop}
\end{figure}

 \begin{figure}[t]\centering
%\vspace*{-4.4cm}
\scalebox{1.2}{\includegraphics{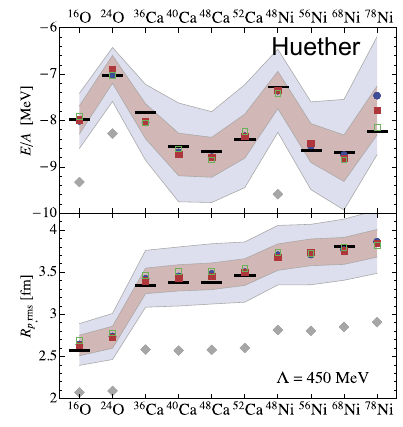}}
%\vspace*{-0.25cm}
\caption{
 Ground-state energies per nucleon (top panel)
 and point-proton rms radii (bottom panel)
 for selected medium-mass isotopes as obtained in the 
 ``H\"uther'' case~\cite{Hut20}.
The light blue and pink bands represent 
the theoretical uncertainties at NNLO and N$^3$LO, respectively.
$\Lambda=450$ MeV.
Black bars indicate the experimental data. (Figure courtesy of R. Roth)}
 \label{fig_roth}
\end{figure}

\begin{figure}[t]\centering
%\vspace*{-4.4cm}
\scalebox{0.7}{\includegraphics{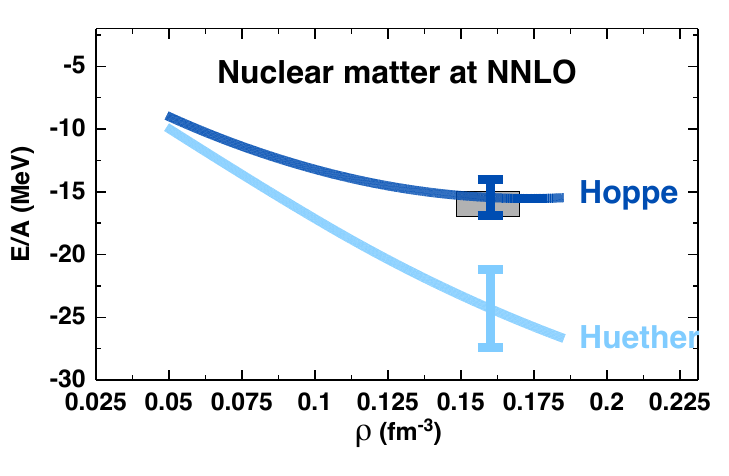}}
%\vspace*{-0.25cm}
\caption{
 Energy per nucleon, $E/A$, as a function of density, $\rho$, of symmetric nuclear matter as obtained in
calculations with the 2NFs and 3NFs consistently at NNLO~\cite{SM20}. 
In the two cases shown, the 2NF is the same, while the 3NFs are the ones used
in the calculations of finite nuclei in the ``Hoppe'' and ``Huether'' cases as denoted.
$\Lambda=450$ MeV in both cases.
The error bars show the theoretical uncertainties around saturation,
which is expected to occur in the area of the gray box. }
 \label{fig_fsd}
\end{figure}

 \begin{itemize}
 \item {\bf ``Magic''}~\cite{Heb11,Heb21}:
 A seemingly successful interaction for the intermediate mass region 
 commonly denoted by ``1.8/2.0(EM)'' (sometimes
 dubbed ``the  Magic force''). It is a
 similarity renormalization group
   (SRG) evolved version of the N$^3$LO 2NF of Ref.~\cite{EM03} complemented by a NNLO 3NF adjusted to the triton binding energy and the 
 point charge radius  of $^4$He. With this force, the ground-state energies all the way up to the
 tin isotopes are reproduced perfectly---but with charge radii being on the smaller side~\cite{Sim17,Mor18}.
 Nuclear matter saturation is also reproduced reasonably well, but at a slightly too high saturation density~\cite{Heb11}.
 
 \item {\bf ``GO''}~\cite{Eks18,Jia20}:
 A family of $\Delta$-full 
NNLO potentials constructed by the G\"oteborg/Oak Ridge (GO) group.
The authors claim to obtain
 ``accurate binding energies and radii for a range of nuclei from $A=16$ to $A=132$,
 and provide accurate equations of state for nuclear matter''~\cite{Jia20}.
 
\item {\bf ``Hoppe''}~\cite{DHS19,Hop19}:
 Recently developed soft chiral 2NFs~\cite{EMN17} at NNLO and N$^3$LO
 complemented with 3NFs at NNLO and N$^3$LO, respectively, to fit the triton binding energy and nuclear matter saturation. These forces applied in
 in-medium similarity renormalization group (IM-SRG~\cite{Her16})
 calculations of finite nuclei up to $^{68}$Ni predict underbinding
and slightly too large radii~\cite{Hop19}, see Fig.~\ref{fig_hop}.

\item {\bf ``H\"uther''}~\cite{Hut20}: The same 2NFs used in ``Hoppe'', but with the 3NFs adjusted
 to the triton and $^{16}$O ground-state energies.  The interactions so obtained reproduce
 accurately experimental energies and point-proton radii of nuclei up to $^{78}$Ni~\cite{Hut20}, see Fig.~\ref{fig_roth}.
 However, when the 2NF plus 3NF combinations of ``H\"uther'' are utilized in nuclear matter, then overbinding and no saturation
 at realistic densities is obtained~\cite{SM20}, see Fig.~\ref{fig_fsd}.
\end{itemize}

\begin{figure}[t]\centering
%\vspace*{-2cm}
\scalebox{1.0}{\includegraphics{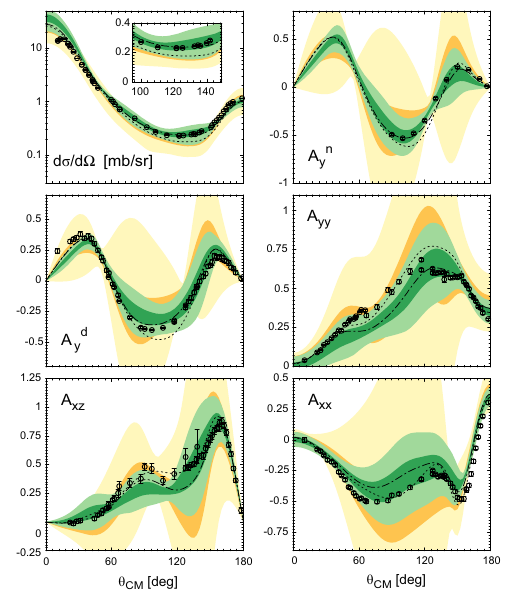}}
%\vspace*{-0.25cm}
\caption{
Predictions for the differential cross section, nucleon and
deuteron analyzing powers $A^n_y$ and $A^d_y$
 as well as deuteron tensor analyzing
powers $A_{yy}$, $A_{xz}$, and $A_{xx}$ in elastic nucleon–deuteron scattering
at a laboratory energy of 135 MeV at NLO (yellow bands) and
NNLO (green bands). 
 The light- (dark-) shaded bands indicate 95\% (68\%)
 confidence levels.
 The dotted (dashed)
lines show the results based on the CD-Bonn $NN$ potential~\cite{Mac01} 
 (CD-Bonn $NN$ potential in combination with the Tucson-Melbourne 3NF~\cite{CH01}).
 Black symbols represent the data together with their experimental errors.
(Reproduced from Ref.~\cite{Epe20}.)}
\label{fig_3N}
\end{figure}

  Obviously, in some cases, there appears to be a problem with achieving simultaneously accurate results
 for nuclear matter and medium-mass nuclei: In the ``Hoppe'' case, nuclear matter is saturated correctly, but nuclei are underbound; while in the
 ``H\"uther'' case, nuclei are bound accurately, but nuclear matter is overbound.
 Other cases seem to have solved this problem.
 But are they all truly {\it ab initio}?
 Our assessment:

 \begin{itemize}
 \item {\bf ``Magic'':}
 The construction of this force includes some inconsistencies.
The 2NF is SRG evolved, while the 3NF is not.
 Moreover, the SRG evolved 2NF
 is used like an original force with the induced 3NFs omitted. 
 Note that {\it ab inito} also implies that the forces are based upon
 some sort of theory in a consistent way. This is here not true
 and, thus, this case is not {\it ab initio}.
 
  \item {\bf ``GO'':}
  In Ref.~\cite{NEM21} it has been shown that the predictions by
  the $\Delta$-full $NN$ potentials at NNLO constructed by the 
G\H{o}teborg-Oak Ridge (GO) group~\cite{Jia20} are up to 40 times outside the 
theoretical error of  chiral EFT at NNLO.
So, they fail on accuracy.
 The reason for their favorable reproduction of the energies (and radii) of intermediate-mass nuclei,
can be traced to incorrect $P$-wave and $\epsilon_1$ mixing parameters~\cite{NEM21}. 
Thus, this case is especially far from being {\it ab initio}.
It is just a repetition of the mistakes of the early 1970s.

\item {\bf ``Hoppe'':}
In this case, the 2NF and 3NF forces are consistently chiral EFT based.
Moreover, the 2NFs are accurate. 
However, there is another accuracy aspect that is, in general, quietly 
ignored~\cite{KVR01,Mar09}: Are the 3NFs accurate?
The accuracy of the chiral 3NF at NNLO was thoroughly investigated in Ref.~\cite{Epe20} 
for a variety of cutoffs ranging from 400-550 MeV 
and large variations of the NNLO 3NF parameters, $c_D$ and $c_E$.
A typical result is shown in Fig.~\ref{fig_3N}.
It is seen that the 3$N$ data are reproduced within the truncation errors
at NNLO (green bands). On the other hand, it is also clearly seen that the theoretical
uncertainties are very large.
Moreover, it was found in Ref.~\cite{Epe20} that the cutoff dependence is weak and that the variations of the 3NF LECs $c_D$ and $c_E$ make only small differences relative to the
large uncertainties.
Thus, we can assume that the NNLO 3NFs used
in ``Hoppe'' will yield results
that lie within the NNLO uncertainties
shown in Fig.~\ref{fig_3N} by the green bands
and, consequently,
the ``Hoppe'' 3NF is accurate.
Hence, ``Hoppe'' passes on all accounts and is, therefore, truly {\it ab initio}.
\item {\bf ``H\"uther'':}
An assessment similar to ``Hoppe'' applies. Thus,  this case is
also truly {\it ab initio}.
 \end{itemize}
 
The bottom line is that not all calculations, which have been published in the 
literature under the label of {\it ab initio}, are really {\it ab initio}.
Indeed, of the cases we considered here, only 50\% pass the test.
But we need to point out that even in the two cases we declared {\it ab initio},
there are concerns. The NNLO predictions by Hoppe and H\"uther
for finite nuclei barely overlap within their theoretical uncertainties and, for nuclear matter, they do not overlap at all.
Obviously, there are problems with the error estimates
and the uncertainties are much larger than the shown ones.
The true NNLO truncation errors of the Hoppe and H\"uther calculations
are probably as large as the differences between the two predictions.
 In this way, the two predictions are actually consistent with
each other, in spite of their seeming discrepancy. Chiral EFT is a
model-independent theory and, thus, different calculations at the same order
should agree within truncation errors.

At N$^3$LO the predictions differ even more. However, 
for current N$^3$LO calculations, a strong caveat is in place.
As pointed out in Ref.~\cite{EKR20}, 
there is a problem with the regularized 3NF at N$^3$LO (and higher orders)
in all present nuclear structure calculations. 
The N$^3$LO 3NFs currently in use are all regularized
by a multiplicative regulator applied to the 3NF expressions that are
derived from dimensional regularization.
This approach leads to a violation of chiral symmetry at N$^3$LO
and destroys the consistency between two- and three-nucleon forces~\cite{EKR20}. Consequently, all current calculations that
include a N$^3$LO 3NF contain an uncontrolled error and are, therefore,
unreliable. When a consistent regularization scheme has been found, the calculations have to be repeated.
At the present time, reliable predictions exist only at
NNLO, NLO, and LO.

 \section{The future: {\it ab initio} plus precision}

 \begin{figure}[t]\centering
%\vspace*{-2cm}
\scalebox{1.0}{\includegraphics{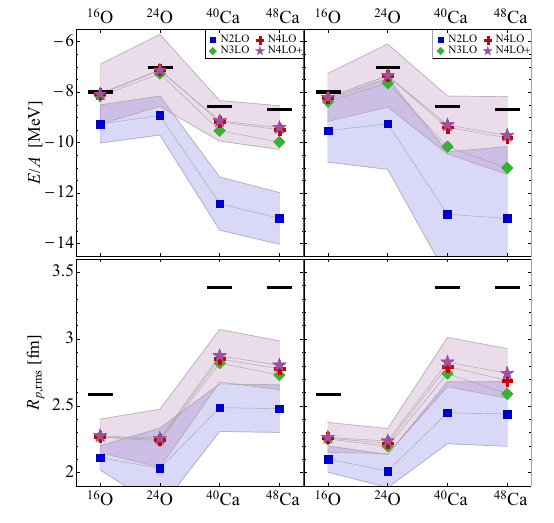}}
%\vspace*{-0.25cm}
\caption{Latest {\it ab initio} predictions by the LENPIC collaboration~\cite{Mar22}:
Ground-state energies and point-proton radii for doubly magic oxygen and
calcium isotopes obtained from the $NN$ potential of Ref.~\cite{RKE18}
complemented by NNLO 3NFs using a cutoff of 450 MeV (left-hand panel)
and of 500 MeV (right-hand panel). The blue squares
represent the predictions by complete NNLO calculations with the 
blue error bands showing the chiral NNLO truncation
uncertainties at the 95\% confidence level.
The green and purple points and pink error bands are based upon
incomplete calculations and are to be ignored.
Black bars indicate the experimental data.
(Reproduced from Ref.~\cite{Mar22} with permission.)
}
\label{fig_LENPIC}
\end{figure}

It is comforting to know that at least a few correct {\it ab initio} 
calculations do exist. But these cases show that the precision at NNLO 
is very poor. The same
is true for the latest LENPIC calculations~\cite{Mar22}, see Fig.~\ref{fig_LENPIC} (which we did not include in our case study,
because nuclear matter results are lacking). 
At N$^3$LO (if one day correct such calculations become available)
the precision will most likely not be substantially better.

As stated at the outset, the purpose of the {\it ab initio} approach
is to test if the tenet of nuclear theory is correct or not.
Within huge errors as, e.~g. in Fig.~\ref{fig_3N}, any approach may
come out right. So, that is not a good basis for a reliable test.
We need more precision!
This is in particular true for the 3NF and the reproduction of the 3$N$ data,
which has been thoroughly investigated in Refs~\cite{Epe20,WGS22}
with the conclusion that, at N$^4$LO, there is a chance to achieve
the desirable precision---for several reasons.
The long- and intermediate-range
topologies of the 3NF at N$^4$LO are expected to be much larger than the corresponding ones at N$^3$LO because, at N$^4$LO, the subleading $\pi NN$ seagull vertex is involved
with LECs $c_i$, which are large~\cite{KGE12,KGE13}. 
This will provide the 3NF at N$^4$LO with more leverage as compared to  N$^3$LO.
Moreover, at N$^4$LO,
13 new 3$N$ contact terms occur~\cite{GKV11} with essentially free parameters introducing considerable flexibility~\cite{Gir19,WGS22} (see also Ref.~\cite{Gir23}).
Worth mentioning is also that,
 at N$^4$LO, the 3NF includes all 20 operators of the most general 3NF~\cite{Epe15}.
 Furthermore, the plentiful N$^4$LO 3NF terms may also provide what is needed
 to improve the status of the medium-mass nuclei and nuclear matter.
 
 Thus, the future of truly microscopic nuclear structure is to go for
 complete N$^4$LO calculations---a gigantic task.

\section{Summary and outlook}

To summarize, let me just reiterate the main statements.

The tenet of microscopic nuclear theory is:

\begin{quote}
{\sf Atomic nuclei
can be accurately described as collections of point-like nucleons interacting via
two- and many-body forces obeying nonrelativistic quantum 
mechanics---the forces being fixed in free-space scattering.}
\end{quote}

And in the {\it ab initio} approach, nuclei are calculated accordingly.

We need to critically
investigate
if the tenet is true. To that end, we have to answer the question:

\begin{quote}
{\it Do the same nuclear forces that explain free-space scattering experiments
also explain the properties of finite nuclei and nuclear matter when
applied in nuclear many-body theory?}
\end{quote}

Either way, the answer is of
fundamental relevance.
The correct answer can only be obtained
if the free-space forces are accurate,
where accurate is defined by:

\begin{quote}
{\it Accurate free-space forces are forces that
predict experiment within the theoretical uncertainty of the applied EFT at the 
given order.}
\end{quote}

Moreover, one would also require that the applied nuclear forces are based
upon some sort of theory in a consistent way.

Without strictly adhering to these principles, the true answer to the fundamental question will not be found.
Once again, the goal is not to obtain ``good'' results, but to understand whether
there are non-negligible medium effects on nuclear forces
when inserted into the nuclear many-body problem.

In our community, the term {\it ab initio} is often used in a way that is too lose and many calculations
that are presented as {\it ab initio} do not pass muster.
Such calculations repeat the mistakes of history and, thus, do not 
move us forward.

The ultimate goal of nuclear theory should be to conduct calculations
that test the tenet with high precision. There is strong evidence that this precision can only be achieved at N$^4$LO of the chiral EFT expansion.
Calculations of this kind, which must also include all many-body forces at that order,
are very challenging, and the current status of {\it ab initio} calculations
is far from meeting that goal. 

In this context, it should be mentioned that the uncertainties of the many-body
calculations must also be included in the error analysis. With calculations
now moving up to heavy nuclei, current many-body techniques need
to be tested critically for which bechmark calculations would be the right tool. 

The work that is left to do in microscopic
nuclear theory is monumental.

\begin{acknowledgements}
This work was supported in part by the U.S. Department of Energy
under Grant No.~DE-FG02-03ER41270.
\end{acknowledgements}

\bibliography{bibRM}

\end{document}